\documentclass[12pt]{iopart}
\usepackage{graphicx}

\begin{document}

\title{What is credible and what is incredible in the measurements of
the Casimir force}

\author{ G.~L.~Klimchitskaya${}^{1,2}$ and
V. M. Mostepanenko${}^{1,3}$
}

\address{${}^1$Department of Physics, Federal University of Para\'{\i}ba,
{\protect \\}
C.P. 5008, CEP 58059--900, Jo\~{a}o Pessoa, Pb-Brazil {\protect \\}
${}^2$North-West Technical University, {\protect \\}
 Millionnaya Street 5, St.Petersburg,
191065, Russia{\protect \\}
${}^3$Noncommercial Partnership
``Scientific Instruments'', {\protect \\}
Tverskaya Street 11, Moscow,
103905, Russia
}


\begin{abstract}
We comment on progress in measurements of the Casimir force and
discuss what is the actual reliability of different experiments.
In this connection a more rigorous approach to the usage of such
concepts as accuracy, precision, and measure of agreement
between experiment and theory, is presented. We demonstrate
that all measurements of the Casimir force employing
spherical lenses with
centimeter-size curvature radii are fundamentally flawed
due to the presence of bubbles and pits on their surfaces.
The commonly used formulation of the proximity force
approximation is shown to be inapplicable for centimeter-size
lenses. New expressions for the Casimir force are derived taking
into account surface imperfections. Uncontrollable deviations
of the Casimir force from the values predicted using the
assumption of perfect sphericity vary by a few tens
of percent within
the separation region from 1 to $3\,\mu$m.
This makes impractical further use of centimeter-size
lenses in experiments on measuring the Casimir force.
\end{abstract}

\section{Introduction}

The preprint \cite{1} reviews some recent experiments on measuring
the Casimir force. The author makes a reservation that he will
consider only ``credible'' experiments. As incredible, the
experiments which have claimed ``1\% or better agreement'' are
meant with a generic reference to review \cite{2}.
In different places of the preprint these unspecified
incredible experiments are characterized as ``1\% level work''
(pp.\ 1, 3), experiments ``that claim 1\% accuracy''
(p.\ 4), and ``experiments claiming 1\% precision'' (p.\ 27).
In the first part of Ref.~\cite{1} the author provides
arguments why it is unclear to him ``what these experiments
really mean''. The second part of Ref.~\cite{1} is largely
devoted to different aspects of author's own work employing
spherical lenses of more than 10\,cm curvature radius.

Below we demonstrate that the author's arguments against
what he
calls a ``1\% level work'' are based on incorrect or
incomplete information. With respect to measurements of
the Casimir force using lenses of centimeter-size
curvature radii,
we show that they are fundamentally flawed. According to
our calculations, experiments of this type may lead to
unpredictable results for the Casimir force, due to
unavoidable deviations from a spherical shape of
mechanically polished and ground surfaces.

The paper is organized as follows. In Sec.~2 we
explain what is incorrect in the argumentation of
Ref.~\cite{1} against precise experiments on measuring
the Casimir force. Here we consider a relationship among
the concepts of accuracy, precision and measure of
agreement with theory, discuss total experimental error,
its constituents, and rules of their combination,
explain the misjudgement of constraints on long-range forces
made in Ref.~\cite{1}. In Sec.~3 it is shown that all
measurements of the Casimir force employing
centimeter-size spherical lenses are fundamentally
flawed. We discuss both the electrostatic calibrations
and the measurement of the Casimir force.
We demonstrate that commonly used simplified form of the
proximity force approximation (PFA) is inapplicable
in the presence of standard imperfections
on the optical surfaces (bubbles and pits) and derive
new expressions for the Casimir force
valid in the presence
of these imperfections. In Sec.~4 some further
objectionable features of Ref.~\cite{1} are discussed.
Section~5 contains our conclusions and discussion.

\section{What is incorrect in the arguments against
precise experiments}

\subsection{Confusion between accuracy, precision and measure of agreement
with theory}

The present state of the art in experiments on measuring the
Casimir force is reflected in the review \cite{2}. Some results
in this review do not necessarily coincide with respective
formulations in original publications because several experiments
were later reanalyzed using more reliable methods of data
processing. At the moment only these updated results of
Ref.~\cite{2} should be used in all discussions.
According to Ref.~\cite{2}, each experiment on measuring the
Casimir force is characterized by a total experimental error,
total theoretical error, and measure of agreement between
experiment and theory determined at some high (usually 95\%)
confidence level. In this respect the above cited
characterizations
of precise experiments in Ref.~\cite{1}, which confuse 1\%
agreement, 1\% accuracy and 1\% precision are completely
misleading. According to Ref.~\cite{2}, the best measure of
agreement between the Casimir pressure measured in the most
precise experiment \cite{3} using a micromachined oscillator
and theory is equal to 1.8\% at the separation $a=400\,$nm
between the test bodies.
As to the best measure of agreement between the measured
Casimir force in sphere-plate geometry and theory in the
most precise experiment \cite{4} using an atomic force
microscope, it is equal to 5.4\% at separations around
$a=80\,$nm \cite{5}.
Hence the ``1\% or better agreement'' is an incorrect
information.

Of even greater concern is the mention of experiments
``that claim 1\% accuracy'' \cite{1}. According to the
review \cite{2}, there are no such experiments. What's
more, measuring of accuracy in percents, as is done in
Ref.~\cite{1} for many times, is in contradiction with
the rigorous understanding of this concept.
According to the International vocabulary of
metrology \cite{6} produced by the Joint Committee for
Guides in Metrology, measurement accuracy is the
closeness of agreement between a measured quantity value
and a true quantity value. As underlined in Ref.~\cite{6},
the concept ``measurement accuracy'' is not a quantity
and is not given a numerical value. This interpretation
is well founded because a true quantity value is in
principle unknown. A measurement is said to be more
accurate when it offers a smaller measurement error.
Thus, it is meaningless to speak about 1\% or any other
numerical degree of accuracy, as is done and erroneously
attributed to some unspecified experiments in Ref.~\cite{1}.

Another concept used in Ref.~\cite{1} in the same context
is the concept of precision. Measurement precision is the
closeness of agreement between measured quantity values
obtained by replicate measurements on the same or similar
objects under specified conditions \cite{6}. Precision
is expressed numerically by measures of imprecision
\cite{6}. Specifically, the experimental errors
 can be used as such measures.
In the experiment \cite{3} the highest measurement
precision is achieved at the separation $a=162\,$nm
where the total relative experimental error is equal to
\cite{2}
\begin{equation}\delta_t\Pi^{\rm expt}=
\frac{\Delta_t\Pi^{\rm expt}}{|\Pi^{\rm expt}|}
=0.19\mbox{\%},
\label{eq0}
\end{equation}
\noindent
where $\Pi^{\rm expt}$ is the measured value of the
physical quantity $\Pi$ (the Casimir pressure between two
parallel plates), and $\Delta_t\Pi^{\rm expt}$
is the total absolute experimental error.
In the experiment \cite{4} the highest precision is
achieved at 63\,nm and corresponds to the total relative
experimental error $\delta_t\Pi^{\rm expt}=1.5$\% \cite{2},
where $\Pi^{\rm expt}$ is the measured Casimir force
between a sphere and a plate.
Although the author of Ref.~\cite{1} writes that
``There is a tendency among workers in this field to
confuse precision with accuracy, of which I am guilty
myself'',
the definitions presented in Ref.~\cite{1} continue to be
unrelated to the rigorous formulations \cite{6}.
Specifically, precision relates not a number of significant
figures provided by a measurement device, as erroneously
stated in Ref.~\cite{1}, but to a closeness between measured
quantity values in replicate measurements (the same
voltmeter, for instance, can be used in different experiments
leading to different precisions).

\subsection{Total experimental error and its constituents}

Reference \cite{1} argues that ``to obtain a given experimental
accuracy, say 1\%, requires that the calibrations and force
measurements must be done to much better than 1\% accuracy...''
However, the suggested arguments that the latter is yet not
possible contain several incorrect statements.
Before we indicate each of the specific mistakes made, let us
emphasize that the word ``accuracy'' used in Ref.~\cite{1}
must be replaced with the word ``precision'' because, as explained
above, accuracy is not given a numerical value and there are no
experiments claiming a 1\% accuracy.

The author of Ref.~\cite{1} is right that if, for instance,
the total experimental error is equal to 1\% all calibration
errors must be smaller accordingly depending on their size and
number. In the list of these errors, however, Ref.~\cite{1}
again confuses by mixing experimental errors,
theoretical errors and agreement between experiment and theory.
Here, we illustrate what are the constituents of the lowest
total experimental error $\delta_t\Pi^{\rm expt}=0.19$\%
at a separation $a=162\,$nm in the most precise experiment
\cite{3}. The relative random error in the Casimir pressure at
$a=162\,$nm  is $\delta_r\Pi^{\rm expt}=0.04$\%
\cite{2}. It is determined from the standard statistical procedure
using Student distribution \cite{7}. The systematic error is
caused by the errors in the measurement of the sphere radius,
of the frequency shift, and of the proximity force approximation
(which is a part of experimental procedure in the indirect
measurement of the Casimir pressure). The resulting relative
systematic error at a separation of 162\,nm is
$\delta_s\Pi^{\rm expt}=0.19$\%.
According to Ref.~\cite{1}, for achieving a 1\% experimental
accuracy (read precision) the sphere radius needs to be measured
to 0.5\% accuracy (precision). Reference~\cite{1} claims that
``the radius measurement is not discussed in sufficient details
in any of papers...'' This is, however, not so. The value of the
sphere radius in the experiment \cite{3} was determined to be
$R=151.2\pm 0.2\,\mu$m leading to the relative error of only
0.13\% \cite{2,3}, i.e., smaller error than is demanded in
Ref.~\cite{1}.
All the details for determination of sphere radius by
means of electrostatic calibrations are provided in
Refs.~\cite{3,8} (specifically, the calibration details are
presented in full in Ref.~\cite{9}).

The other sources of errors considered in Ref.~\cite{1} are
unrelated to the experimental precision. Thus, the knowledge
of the optical properties of the surfaces is not needed for
the determination of precision. The discussion of errors in
the Casimir force induced by the errors in absolute separations
bears no relation to force and pressure measurements as well.
The separation distance is an independent quantity and is
measured with its own measurement error (in Ref.~\cite{3}
the latter is equal to 0.6\,nm).
Both these errors are important for the comparison
between experiment and theory, but have nothing to do with
the achieved experimental precision of force and pressure
measurements reviewed in Ref.~\cite{2}.

\subsection{Is it really uncertain how to combine different errors
and uncertainties?}

As discussed above, the total experimental error results from
the combination of random and systematic errors. In its turn,
the systematic error has several constituents.
According to Ref.~\cite{1}, precision measurement experts still
debate whether these errors and uncertainties can be added
in quadrature or be simply added.
Regarding this statement we suggest that the author of Ref.~\cite{1}
was guided by outdated information.
It is common knowledge that errors and uncertainties are random
quantities and are characterized by some distributions \cite{10}.
The composition law of several random quantities depends on
the specific form of these distributions.
In the measurement of the Casimir force it is usually supposed
that all systematic errors in the form of systematic deviations
(i.e., biases in a measurement which always make the measured
value higher or lower than the true value) are already removed
using some known process, i.e., through a calibration.
The remaining systematic errors are the errors of a calibration
device and have the meaning of the smallest fractional devision
of the scale of the device. Such systematic errors are random
quantities characterized by a uniform distribution (equal
probability). The errors in an approximate theoretical formula
used to convert a directly measured quantity into an indirectly
measured one are also distributed uniformly.
Then the resulting systematic error at a chosen confidence
level $\beta$ is obtained from its constituents
$\Delta_s^{\!(i)}\Pi^{\rm expt}$ ($i=1,2,\,\ldots ,J$) using the
following statistical rule \cite{10}
\begin{equation}
\Delta_s\Pi^{\rm expt}=\min\left[
\sum_{i=1}^{J}\Delta_s^{\!(i)}\Pi^{\rm expt},
k_{\beta}^{(J)}\sqrt{
\sum_{i=1}^{J}(\Delta_s^{\!(i)}\Pi^{\rm expt})^2\,\,\,}\right].
\label{eq1}
\end{equation}
\noindent
Here, $k_{\beta}^{(J)}$ is a tabulated coefficient. The above
value of $\delta_s\Pi^{\rm expt}=0.19$\% at $a=162\,$nm in the
experiment \cite{3} (see Sec.~2.2) was obtained using Eq.~(\ref{eq1})
with $J=3$, $\beta=0.95$, and $k_{0.95}^{(3)}=1.1$ \cite{2,5}.

Contrary to Ref.~\cite{1}, statistical rules for the combination
of random and systematic errors have also been much studied.
The random error is described by the normal or Student
distribution. The resulting systematic error is described by
a combination of uniform distributions. It can be shown that
if the resulting systematic error is also assumed to be
distributed uniformly, the total experimental error will be
overestimated. Thus, this assumption is conservative and can be
used safely. There are several methods in statistcs how to
combine errors described by normal and uniform distributions
\cite{10}. A widely used method puts
\begin{equation}
\Delta_t\Pi^{\rm expt}=\Delta_r\Pi^{\rm expt}, \qquad
\Delta_t\Pi^{\rm expt}=\Delta_s\Pi^{\rm expt},
\label{eq2}
\end{equation}
\noindent
or
\begin{equation}
\Delta_t\Pi^{\rm expt}=q_{\beta}(r)\left[\Delta_r\Pi^{\rm expt}
+\Delta_s\Pi^{\rm expt}\right]
\label{eq3}
\end{equation}
\noindent
depending on what respective inequality is fulfiled for all $a$
over the entire measurement range
\begin{equation}
r(a)<0.8, \qquad r(a)>8, \qquad
\mbox{or}\qquad
0.8\leq r(a)\leq 8.
\label{eq4}
\end{equation}
\noindent
Here, the quantity $r(a)$ is defined as
\begin{equation}
r(a)=\frac{\Delta_s\Pi^{\rm expt}(a)}{s_{\bar\Pi}(a)},
\label{eq4a}
\end{equation}
\noindent
where $s_{\bar\Pi}(a)$ is the variance of the mean
of a measured quantity $\Pi^{\rm expt}$.
The coefficient $q_{\beta}(r)$ at a confidence level
$\beta=0.95$ varies between 0.71 and 0.81 depending on the
value of $r(a)$. Note that the value of the relative total error
in the experiment \cite{3} at $a=162\,$nm was obtained using the
second equality in Eq.~(\ref{eq2}).
We emphasize that the dominance of the resulting systematic error
over the random error within the entire measurement range
achieved in the experiment of Ref.~\cite{3} is the distinguishing
feature of precise experiments of a metrological quality.

{}From the above facts one can conclude that precision measurement
experts have gone far beyond debates whether uncertainties can be added
in quadrature or simply added.

\subsection{Misjudgement of constraints on long-range forces following
from the most precise Casimir experiment}

The measure of agreement between the Casimir pressures measured in
the most precise experiment \cite{3} and calculated theoretically
was used to obtain the strongest constraints on the parameters of
long-range Yukawa-type forces in the interaction range of several
tens of nanometers \cite{3,5}. In doing so the Yukawa pressure was
calculated \cite{3} by the application of the PFA.
Reference \cite{1} informs the reader that the use of this
approximation has been criticized in Ref.~\cite{11}.
The author of Ref.~\cite{1} repeats the conclusion of
Ref.~\cite{11} that the PFA ``only applies to a force that depends
on the location of body surfaces'' and ``is not valid for
the volume integral required for calculating the anomalous
force''.

This conclusion is, however, incorrect as is demonstrated in
available literature overlooked by the author of Ref.~\cite{1}.
In Ref.~\cite{12} it is shown that the PFA is applicable
for the calculation of the Yukawa force under conditions
that the separation $a$ and interaction range $\lambda$ are much
smaller than the sphere radius $R$ and the plate thickness $D$.
All these conditions are satisfied with a large safety margin
in the experimental setup of Ref.~\cite{3}. In Ref.~\cite{13}
the respective Yukawa pressure in the setup of Ref.~\cite{3}
was calculated both exactly and using the PFA with coinciding
results. The purported ``corrections'' to the calculation of
Ref.~\cite{12} pointed out in Ref.~\cite{11} were shown to
be invalid and based on a simple misunderstanding \cite{13}.
What's more, one of the authors of Ref.~\cite{11} (R.O.)
recognized \cite{14} that the issue raised in their paper
``is not of practical concern for current experiments''.
Thus, the reliability of constraints on the parameters
of Yukawa interactions obtained in Ref.~\cite{3} is
beyond doubt. However, the author of Ref.~\cite{1}
included only an incorrect reference to the paper \cite{12}
(the title is taken from one paper and the publication
data from another; see Ref.~[24] in \cite{1}) in his list
of references. As to important Refs.~\cite{13,14},
where the validity of constraints of Ref.~\cite{3} is
reinforced in an unambiguous way, the author of
Ref.~\cite{1} didnot mention them.

\section{Why Casimir force measurements using centimeter-size
spherical lenses are fundamentally flawed}

\subsection{Anomalies in electrostatic calibrations}

Observations of anomalous electrostatic forces in the lens-plate
geometry for lenses of centimeter-size curvature radii is the
subject of wide speculation (see, e.g., Refs.~\cite{15,16,17}).
In Ref.~\cite{18} it was shown that anomalous behavior of the
electrostatic force can be explained due to deviations
of the mechanically polished and ground surfaces of
centimeter-size lenses from a perfect spherical form.
The point is that the typical surface of a centimeter-size lens
is characterized in terms of scratch and dig optical
surface specification data. In particular, depending on the
quality of lens used, bubbles and pits with a diameter
varying from $30\,\mu$m to 1.2\,mm are allowed on the
surface \cite{19}. There may be also scratches with a width
varying from 3 to $120\,\mu$m \cite{19}. The problem of
bubbles on the centimeter-size lens surface should not be
reduced to the fact that lens curvature radius $R$ is
determined with some error. The thickness of each bubble should
of course be less than the absolute error in the measurement of lens
curvature radius (for a lens with $R=15.10\,$cm in Ref.~\cite{16},
for instance, $\Delta R=0.05\,$cm). The crucial point is that
curvature radii of bubbles can be orders of magnitude different,
as compared to $R$. This allows one to suggest models leading to
quite different (``anomalous'') dependence of electrostatic force
on separation in comparison with the case of perfect spherical
surfaces \cite{18}.

Reference \cite{1} mentions the possibility that the anomalous
electrostatic forces are due to simple geometrical effects
without reference to the source of this idea (Ref.~\cite{18} is
missing in the list of references in \cite{1}).
According to Ref.~\cite{1}, this possibility ``is credibly
discarded''
in Ref.~\cite{20}. The author of Ref.~\cite{1} does not inform the
reader that computations of Ref.~\cite{20} were repeated in e-print
\cite{21} and shown to be not reproducible.
Thus, there is no scientific objection against the possibility
that anomalous electrostatic forces are due to deviations of
mechanically polished and ground surfaces from perfect sphericity.
Furthermore, some of the authors of Ref.~\cite{20}
(D.A.R.D.\ and R.O.) recently recognized \cite{22} that local
geometrical deformations of the surface can really lead to an
anomalous electrostatic force not only in sphere-plate
geometry, but for a cylindrical lens in close proximity to
the plate as well. According to Ref.~\cite{22},
``this is certainly a crucial point to be taken into account
in future experiments''. This reference, however, is missing
in the list of references in \cite{1}.
An extensive consideration of the electrostatic calibrations
in Ref.~\cite{1} always assumes perfect sphericity of the
lens surface.

Another misrepresented point directly relevant to electrostatic
calibrations is the dependence of the contact potential on the
separation distance. According to Ref.~\cite{1}, every paper
on the Casimir effect ``that has bothered measuring the contact
potential as a function of distance has shown an apparent
distance dependence of that potential''. This is, however, not
the case. In Ref.~\cite{3} the contact potential was carefully
measured as a function of separation and found to be constant.
The respective measurement data of the electrostatic
calibrations are published in Refs.~\cite{9,18}. Constant
contact potential was observed in all other experiments by
R.\ S.\ Decca as well (review of these experiments can be found
in Refs.~\cite{2,5}). Independent on separation contact
potential was also reported in Refs.~\cite{4,23,24,25}
and in all other experiments by U.\ Mohideen (see
Refs.~\cite{2,5} for a review). It is notable that
all these experiments were performed in high vacuum with small
spheres of order $100\,\mu$m curvature radii.

\subsection{Influence of surface imperfections on the Casimir force
for lenses of centimeter-size curvature radius}

The Casimir force is far more sensitive than the electrostatic
force to the bubbles and pits that are unavoidably present
on the mechanically polished and ground surface of any lens
of centimeter-size curvature radius. The physical reason is that
the Casimir force falls with the increase of separation distance
more rapidly than the electric force.
As a result, it is determined by  smaller regions near the points
of closest approach of the surfaces. If the local curvature
radius on the lens surface near the point of closest approach to
the plate is significantly different from the mean lens
curvature radius $R$, the impact on the Casimir force can be
tremendous. Below we demonstrate that just this happens due to
the presence of bubbles and pits on a lens surface.
For the sake of simplicity, we consider ideal metal surfaces.
However, it is easily seen that all conclusions obtained are
preserved for real bodies as well.

The Casimir force in sphere-plate geometry under the experimental
conditions $a\ll R$ is usually calculated using the PFA \cite{2,5}.
According to the most general formulation of the PFA \cite{27},
the unknown force between the elements of curved surfaces is
approximately replaced with a known force per unit area of the plane
surfaces (i.e., a pressure) at the respective separation
multiplied by an area element.
Applied to a spherical lens of thickness $D$ above a plane
$z=0$, the PFA represents the force between them in the form
\begin{equation}
F_{sp}(a,T)=\int_{\Sigma}d\sigma P(z,T).
\label{eq5}
\end{equation}
\noindent
Here, $d\sigma$ is the element of plate area, $\Sigma$ is the
projection of the lens onto the plate, $a$ is the shortest
separation between them, $z=z(x,y)$ is the equation of a lens
surface, and $P(z,T)$ is the pressure for two
plane parallel plates at a separation $z$ at temperature $T$.

We choose the origin of a cylindrical coordinate system on
the plane $z=0$ under the lens center. Then for a perfectly
shaped spherical lens the coordinate $z$ of any point
of its surface is given by
\begin{equation}
z=R+a-(R^2-\rho^2)^{1/2}, \quad
\rho^2=x^2+y^2.
\label{eq6}
\end{equation}
\noindent
In this case Eq.~(\ref{eq5}) leads to
\begin{equation}
\!\!\!\!\!\!\!\!\!\!\!\!\!\!
F_{sp}^{\rm perf}(a,T)=2\pi\int_{0}^{\sqrt{2RD-D^2}}\!\!\!\rho d\rho P(z,T)=
2\pi\int_{a}^{D+a}(R+a-z)P(z,T)dz.
\label{eq7}
\end{equation}
\noindent
Keeping in mind that the Casimir pressure is expressed as
\begin{equation}
P(z,T)=-\frac{\partial{\cal F}_{pp}(z,T)}{\partial z},
\label{eq8}
\end{equation}
\noindent
where ${\cal F}_{pp}(z,T)$ is the free energy per unit area of parallel
plates, and integrating by parts in Eq.~(\ref{eq7}), one arrives
at
\begin{equation}
\!\!\!\!\!\!\!\!\!\!\!\!\!\!\!\!\!\!\!\!\!\!\!\!\!\!\!\!\!\!\!
F_{sp}^{\rm perf}(a,T)=2\pi R{\cal F}_{pp}(a,T)-2\pi(R-D){\cal F}_{pp}(D+a,T)-
2\pi\int_{a}^{D+a}\!\!{\cal F}_{pp}(z,T)dz.
\label{eq9}
\end{equation}
\noindent
We consider centimeter-size spherical lenses satisfying
a condition $a\ll D$. For such lenses
${\cal F}_{pp}(D+a,T)\ll{\cal F}_{pp}(a,T)$.
Because of this, one can neglect the second term on the right-hand
side of Eq.~(\ref{eq9}) in comparison with the first \cite{28}.
It can be also shown \cite{28,29} that the first term on the right-hand
side of Eq.~(\ref{eq9}) is in excess of the third by a factor of $R/a$.
This allows one to neglect the third term and arrive to what is
called the simplified formulation of the PFA \cite{28,29}
\begin{equation}
F_{sp}^{\rm perf}(a,T)\approx 2\pi R{\cal F}_{pp}(a,T)
\label{eq10}
\end{equation}
\noindent
widely used for both spherical lenses and for spheres [note that for
a semisphere the second term on the right-hand
side of Eq.~(\ref{eq9}) is identically equal to zero].

For two parallel ideal metal plates spaced $z$ apart the Casimir
free energy per unit area is given by \cite{5,26}
\begin{equation}
{\cal F}_{pp}(z,T)=\frac{k_BT}{\pi}\sum_{l=0}^{\infty}
{\vphantom{\sum}}^{\!\prime}\int_{0}^{\infty}k_{\bot}dk_{\bot}
\ln(1-e^{-2zq_l}).
\label{eq11}
\end{equation}
\noindent
Here, $k_B$ is the Boltzmann constant, $k_{\bot}$ is the magnitude
of the projection of the wave vector on the plates,
$q_l^2=k_{\bot}^2+\xi_l^2/c^2$, $\xi_l=2\pi k_BTl/\hbar$ with
$l=0,\,1,\,2,\,\ldots$ are the Matsubara frequencies, and the
primed summation sign means that the term with $l=0$ is
multiplied by 1/2. For the sake of convenience in computations,
we rewrite Eq.~(\ref{eq11}) in terms of a dimensionless
integration variable $y=2aq_l$ and expand the logarithm in power
series
\begin{equation}
{\cal F}_{pp}(z,T)=-\frac{k_BT}{4\pi z^2}\sum_{l=0}^{\infty}
{\vphantom{\sum}}^{\!\prime}\int_{\tau_z l}^{\infty}ydy
\sum_{n=1}^{\infty}\frac{e^{-ny}}{n}.
\label{eq12}
\end{equation}
\noindent
Here, the dimensionless parameter $\tau_z$ is defined as
$\tau_z=4\pi zk_BT/(\hbar c)$. After performing integration
and then the summation with respect to $l$, the following
result is obtained:
\begin{equation}
{\cal F}_{pp}(z,T)=-\frac{k_BT}{4\pi z^2}\left[
\frac{\zeta(3)}{2}+
\sum_{n=1}^{\infty}\frac{e^{-\tau_zn}}{n^2(1-e^{-\tau_zn})}
\left(\frac{1}{n}+\frac{\tau_z}{1-e^{-\tau_zn}}\right)\right],
\label{eq13}
\end{equation}
\noindent
where $\zeta(x)$ is the Riemann zeta function. Note that the
first contribution on the right-hand side of Eq.~(\ref{eq13})
coincides with the high temperature limit of the free energy.
This is quite reasonable if to take into account that
$\tau_z=2\pi T/T_{\rm eff}$, where the effective temperature
is defined from $k_BT_{\rm eff}=\hbar c/(2z)$.

Now we are in a position to compute the Casimir force between real
spherical lens of large curvature radius with bubbles and pits
of different types and a plane plate. It is common to use the
simplified formulation of the PFA (\ref{eq10}) in sphere-plate
geometry for both small spheres of about $100\,\mu$m radii
and large spherical lenses (see, for instance,
Refs.~\cite{3,4,16,30}). In doing so the role of bubbles and
pits on the surface of lenses of centimeter-size curvature
radii is simply disregarded. Equation (\ref{eq10}), however,
is not applicable for real lenses with large curvature radii
because it assumes perfect spherical surface.
For such lenses one should use a general formulation of the
PFA in Eq.~(\ref{eq5}). To illustrate this fact, we perform
calculations for three typical model imperfections on the
spherical surface near the point of closest approach to the
plate allowed by the optical surface specification data
\cite{19}.

\begin{figure*}[t]
\vspace*{-18.5cm}
\hspace*{-2.5cm}\includegraphics{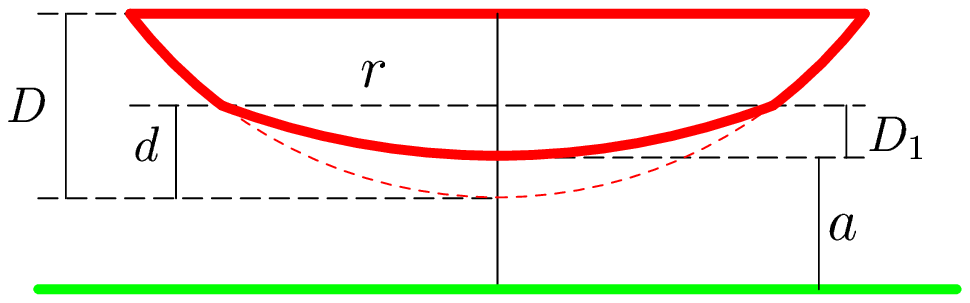}
\vspace*{-8.cm}
\caption{
The configuration of a spherical lens with curvature radius $R$
possessing a surface imperfection at the point of closest
approach to a plate. The bubble curvature radius is $R_1>R$.
The relative sizes of the lens and imperfection are shown
not to scale.
}
\end{figure*}
As the first example, we consider a bubble of the curvature radius
$R_1=25\,$cm which is larger than the curvature radius $R=15\,$cm
of the lens used (see Fig.~1). The thickness of the spherical
lens formed by the bubble is chosen to be $D_1=0.5\,\mu$m
(this is much less than typical absolute error
$\Delta R=0.05\,$cm in the measurement of centimeter-size
lenses curvature radius). The radius of the bubble is
determined from $r^2=2R_1D_1-D_1^2\approx 0.25\,\mbox{mm}^2$,
leading to $2r=1\,\mbox{mm}<1.2\,$mm, i.e., less than
a maximum value allowed by the optical surface specification data
\cite{19}. Respective quantity $d$ defined in Fig.~1 is equal to
$d\approx r^2/(2R)\approx 0.83\,\mu$m. Then the flattening
of a lens surface at the point of closest approach to the
plate is $d-D_1\approx 0.33\,\mu$m which is much less than
$\Delta R$.

The general formulation of the PFA (\ref{eq5}) should be applied
taking into account that the surface of the bubble is described
by the equation
\begin{equation}
z=R_1+a-(R_1^2-\rho^2)^{1/2},
\label{eq14}
\end{equation}
\noindent
where $a$ is the distance between the bottom point of the bubble
and the plate (see Fig.~1). In this notation the surface of the
lens is described by the equation
\begin{equation}
z=R+D_1-d+a-(R^2-\rho^2)^{1/2}.
\label{eq14a}
\end{equation}
\noindent
Using Eqs.~(\ref{eq14}) and (\ref{eq14a}) one arrives, instead
of Eq.~(\ref{eq7}), at
\begin{eqnarray}
F_{sp}(a,T)&=&2\pi\int_{a+D_1}^{a+D_1-d+D}\!\!\!
(R-z+D_1-d+a)P(z,T)dz
\nonumber \\
&&+
2\pi\int_{a}^{a+D_1}(R_1-z+a)P(z,T)dz.
\label{eq15}
\end{eqnarray}
\noindent
Now we take into consideration that the quantities
$a$, $d$, and $D_1$ are smaller than the error in the
determination of large radii $R$ and $R_1$.
Then one can rearrange Eq.~(\ref{eq15}) to the form
\begin{equation}
\!\!\!\!\!\!\!\!\!\!
F_{sp}(a,T)\approx 2\pi\int_{a+D_1}^{a+D_1-d+D}\!\!\!
(R-z)P(z,T)dz
+2\pi R_1\int_{a}^{a+D_1}P(z,T)dz.
\label{eq16}
\end{equation}
\noindent
Here, the first integral on the right-hand side is calculated
similar to Eqs.~(\ref{eq7}) and (\ref{eq9}) leading to
$2\pi R{\cal F}_{pp}(a+D_1,T)$. Calculating the second integral
with the help of Eq.~(\ref{eq8}), one finally obtains
\begin{equation}
F_{sp}(a,T)\approx 2\pi (R-R_1){\cal F}_{pp}(a+D_1,T)+
2\pi R_1{\cal F}_{pp}(a,T).
\label{eq17}
\end{equation}

\begin{figure*}[t]
\vspace*{-15.6cm}
\hspace*{-1.cm}\includegraphics{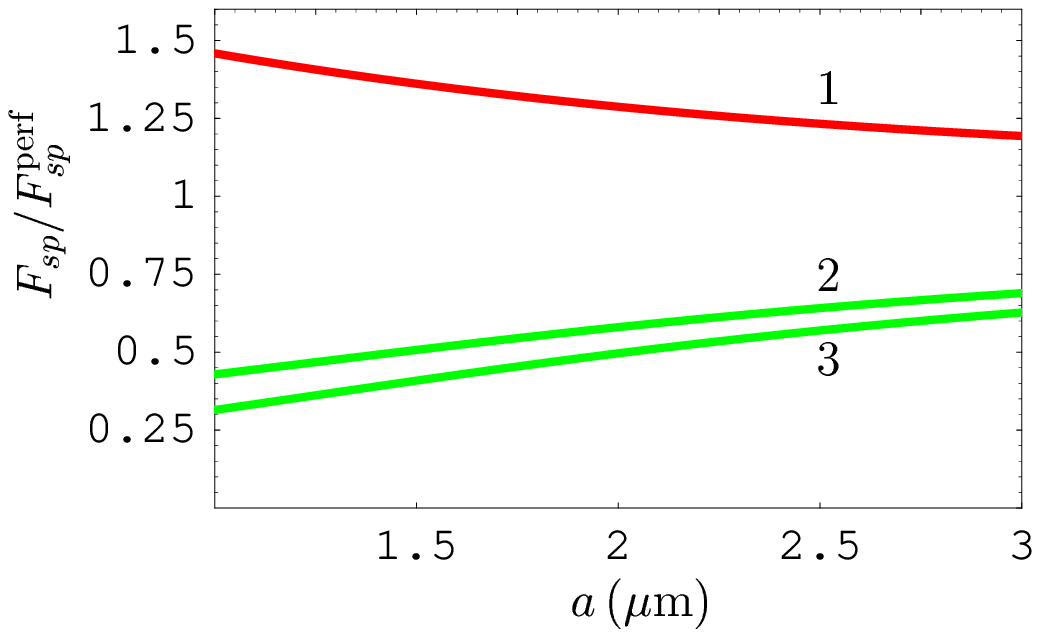}
\vspace*{-7.cm}
\caption{
The normalized Casimir force acting between a sphere with
surface imperfections of different types and a plate as
a function of separation. Lines 1, 2, and 3 are for the
surface inperfections shown in Figs.~1 and 3(a,b),
respectively.
}
\end{figure*}
We present a few computational results demonstrating that the
Casimir force in Eq.~(\ref{eq17}) taking the flattening of a
lens surface into account deviates significantly from the Casimir
force $F_{sp}^{\rm perf}$ in Eq.~(\ref{eq10}) obtained for
perfect spherical surface. Computations of the quantity
$F_{sp}(a,T)/F_{sp}^{\rm perf}(a,T)$ were performed using
Eq.~(\ref{eq13}) at $T=300\,$K within the separation region
from 1 to $3\,\mu$m (see the line labeled 1 in Fig.~2).
As can be seen in Fig.~2, in the presence of a bubble leading to
a flattening of lens surface shown in Fig.~1, the use of
Eq.~(\ref{eq10}) for perfect spherical surface instead of
Eq.~(\ref{eq17}) considerably underestimates the magnitude of
the Casimir force. Thus, at separations $a=1.0$, 1.5, 2.0,
2.5, and $3.0\,\mu$m the quantity $F_{sp}/F_{sp}^{\rm perf}$
is equal to 1.458, 1.361, 1.287, 1.233, and 1.193,
respectively, i.e., the underestimation varies from 46\% at
$a=1\,\mu$m to 19\% at $a=3\,\mu$m.

\begin{figure*}[b]
\vspace*{-11.cm}
\hspace*{-5.cm}\includegraphics{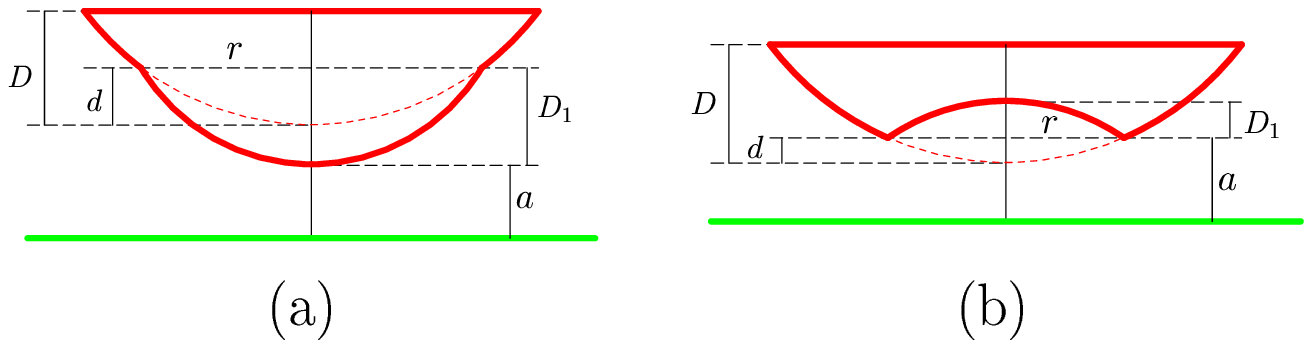}
\vspace*{-15.5cm}
\caption{
The configuration of a spherical lens with curvature radius $R$
possessing a surface imperfection at the point of closest
approach to a plate. (a) The bubble curvature radius is $R_1<R$.
(b) The pit curvature radius is $R_1<R$.
The relative sizes of the lens and imperfection are shown
not to scale.
}
\end{figure*}
Now we consider two more examples of surface imperfection,
specifically, a bubble with the curvature radius $R_1=5\,$cm
[see Fig.~3(a)] and a pit with the curvature radius $R_1=12\,$cm
[see Fig.~3(b)]. In both cases the curvature radius of the
lens remains the same $R=15\,$cm. For the bubble we choose
$D_1=1\,\mu$m which results in $r\approx 0.32\,$mm,
$d\approx 0.33\,\mu$m, and $D_1-d\approx 0.67\,\mu$m
in agreement with allowed values.
Equation (\ref{eq17}) is evidently preserved with new values
of parameters. The computed values of the quantity
$F_{sp}(a,T)/F_{sp}^{\rm perf}(a,T)$ as a function of separation
are shown by the line labeled 2 in Fig.~2. It can be seen that
in this case the assumption of perfect sphericity of a lens
surface considerably overestimates the magnitude of the
Casimir force. Thus, at separations $a=1.0$, 1.5, 2.0,
2.5, and $3.0\,\mu$m the values of the
quantity $F_{sp}/F_{sp}^{\rm perf}$
are equal to 0.429, 0.507, 0.580, 0.641, and 0.689,
respectively, i.e., overestimation varies from 57\% at
$a=1\,\mu$m to 36\% at $a=3\,\mu$m.

Now we deal with a pit shown in Fig.~3(b). Here, the lens surface
near the point of closest approach to the plate is concave up,
i.e., in the direction of lens center. The related parameters are
$D_1=1\,\mu$m,  $r\approx 0.49\,$mm,
$d\approx 0.8\,\mu$m, and $d+D_1\approx 1.8\,\mu$m.
The pit surface is described by the equation
\begin{equation}
z=a+D_1-R_1+(R_1^2-\rho^2)^{1/2}.
\label{eq18}
\end{equation}
\noindent
Here, $a$ is the separation distance between the plate and points
of a circle on the lens surface closest to it.
The surface of the
lens is described as
\begin{equation}
z=R+a-d-(R^2-\rho^2)^{1/2}.
\label{eq19}
\end{equation}
\noindent
Repeating calculations that have led to Eq.~(\ref{eq17})
with the help of Eqs.~(\ref{eq18}) and (\ref{eq19}),
we obtain
\begin{equation}
F_{sp}(a,T)\approx 2\pi (R-R_1){\cal F}_{pp}(a,T)+
2\pi R_1{\cal F}_{pp}(a+D_1,T).
\label{eq20}
\end{equation}

The computational results using Eqs.~(\ref{eq10}), (\ref{eq13})
and (\ref{eq20}) are shown by the line labeled 3 in Fig.~2.
Once again, the assumption of perfect lens sphericity
significantly overestimates the magnitude of the
Casimir force. Thus, at separations $a=1.0$, 1.5, 2.0,
2.5, and $3.0\,\mu$m the
ratio $F_{sp}/F_{sp}^{\rm perf}$
is equal to 0.314, 0.409, 0.496, 0.570, and 0.627,
respectively, i.e., overestimation varies from 69\% at
$a=1\,\mu$m to 37\% at $a=3\,\mu$m.

To conclude this section, we have shown that depending on the
character of imperfections on the lens surface near the
points of closest approach to the plate, the use of the PFA
in the simplest form (\ref{eq10}) can lead to either
underestimated or overestimated Casimir force by a few tens of
percent. Keeping in mind that the exact position of the point
of closest approach on a spherical surface cannot be controlled
with sufficient precision, it seems impossible to determine
the character of surface imperfections near the
point of closest approach microscopically for subsequent
use of Eqs.~(\ref{eq17}) or (\ref{eq20}). This leads us
to the conclusion that measurements of the Casimir force
by using spherical lenses of centimeter-size
curvature radii are
fundamentally flawed in the sense that they can lead to
unpredictable measurement results which cannot be reliably
compared with theory.

\section{Further objectionable features}

\subsection{Whether or not the Casimir force is simply the
retarded van der Waals force?}

Reference \cite{1} gives negative answer to this question
because ``the Casimir force does not depend on the properties of
the individual atoms of the plates, but on their bulk
properties''.
The same is, however, correct for the van der Waals force.
It is common knowledge that the Lifshitz theory expresses
both the van der Waals and Casimir forces in terms of the
dielectric permittivity of plate material which is a bulk
property. The Lifshitz theory presents the unified
description of the van der Waals and Casimir forces for
both dielectric and metallic plates. According to this
theory, the van der Waals force occurs in the nonretarded
limit. With the increase of separation after some transition
regime, the van der Waals force transforms into the Casimir force
in the retarded limit. Reference \cite{1} argues that
``if the Casimir force was simply the retarded van der Waals
force it would make little sense consider modifying the
Casimir force, in a fundamental way, by altering the mode
structure imposed by specially tailored boundary
conditions''. This argument, however, does not work
because in the classical Ref.~\cite{31} the nonretarded
van der Waals force was just obtained from the mode structure
determined by boundary conditions. Because of this,
in the configuration of two material bodies
separated with a gap
it is beyond reason to make difference between the
Casimir force and retarded van der Waals force.

\subsection{Is there conflict between the Casimir effect and
electrical engineering?}

Reference \cite{1} applies classical Maxwell equations in
combination with the plasma model at low frequency and
arrives at an effect that is not experimentally observed.
Basing on this, it is concluded that ``we are faced with
discarding over a century of electrical engineering
knowledge in order to explain a few 1\% level Casimir
force of questionable accuracy''. This conclusion is,
however, unjustified. Electrical engineering deals with
real electromagnetic fields. It is a matter of common
 knowledge that
classical Maxwell equations in the  quasistatic limit lead to
the Drude model dielectric permittivity which is inverse
proportional to the frequency, and not to the plasma
model which is only applicable in the range of infrared
frequencies. This fact is underlined (see, for instance,
Refs.~\cite{2,5,32}) when the plasma model is used for
the theoretical description of the Casimir force.
The delicate point, overlooked in Ref.~\cite{1},
is that both classical electrodynamics and
electrical engineering deal with
real electromagnetic fields, whereas the Casimir effect
deals with fluctuating electromagnetic fields possessing
zero mean value. One of the postulates of quantum
statistical physics that physical system reacts in the
same way on real and fluctuating electromagnetic fields
is presently under question from both theoretical and
experimental sides. This does not touch any fact related
to real electromagnetic fields. In view of this, it
seems baseless to write about discarding over a centure
of electrical engineering knowledge.

\subsection{Technical mistakes}

Many references, equations, and formulations in Ref.~\cite{1}
are incorrect. Below we indicate only a few examples.
In Sec.~2.4 we already wrote that Ref.~[24] in Ref.~\cite{1}
is incorrect. Another example is Ref.~[7] in Ref.~\cite{1}.
According to it, there is a Comment by three authors
on Lamoreaux's paper \cite{30}. In fact there was a
Comment \cite{33} by the two authors and
Lamoreaux's reply \cite{34}.

Equation (5) in Ref.~\cite{1} for the Matsubara frequencies
is incorrect because of missing multiple 2 on the right-hand
side.

In Eq.~(6) of Ref.~\cite{1} the velocity of light $c$ in
the denominator on the right-hand side must be deleted
because the electrical conductivity $\sigma$ is measured
in $\mbox{s}^{-1}$ [the author uses CGSE units, where the
dielectric permittivity expressed by his Eq.~(6) must be
dimensionless]. For the same reason $c$ in
the denominator on the right-hand side of Eq.~(8)
in Ref.~\cite{1} must be
replaced with $c^2$.

According to the explanations below Eq.~(7) of Ref.~\cite{1},
which introduces the generalized plasma model,
``$\varepsilon$ is the usual Drude model permittivity,
for example...'' This is, however, incorrect.
Here, $\varepsilon$ is the dielectric permittivity
describing the interband transitions of core electrons
with nonzero oscillator frequencies. Thus, the
Drude-like term is excluded. Then, according to Ref.~\cite{1},
Eq.~(7) ``is assumed to be valid at very high frequencies,
much above the resonances in the system of atoms and charges
that comprise the plates''. This is also incorrect.
For real electromagnetic fields the generalized plasma
model is valid in the region of infrared optics, which is
below the resonances describing interband transitions,
and in the region of these resonances as well.

According to Ref.~\cite{1}, ``Until now, no significant
or non-trivial corrections to the Casimir force due to
boundary modifications have been observed
experimentally.'' Concerning the work \cite{35} on
the Casimir force between a Au coated sphere and a Si
plate structured with rectangular corrugations, it is
recognized that it presents ``a convincing measurement
of a non-trivial geometrical influence on the Casimir
force'' (the reference to this work in Ref.~\cite{1}
contains mistakes). As was noted in Ref.~\cite{1},
however, the calculations in Ref.~\cite{35}
``were not for real materials.''
Contrary to what is stated in Ref.~\cite{1}, there is
a work \cite{25}, where nontrivial geometrical effects
were observed and found to be in agreement with exact
scattering theory in the configuration of a
sinusoidally corrugated sphere above a sinusoidally
corrugated plate. In so doing real material properties
were taken into account and computations were done at
the laboratory temperature $T=300\,$K.
This work was not mentioned
in Ref.~\cite{1}.

\section{Conclusions and discussion}

In the foregoing we have drawn attention to a few facts
relevant to the progress in measurements of the Casimir
force. We have discussed what is incorrect in the
argumentation of Ref.~\cite{1} against precise experiments
performed up to date and clarified some terminology
of metrological character.

The main new result of this paper is that all measurements
of the Casimir force with centimeter-size spherical lenses are
fundamentally flawed due to unavoidable deviations from
a spherical shape arising from their manufacture.
We have demonstrated that bubbles and pits satisfying
constraints imposed by the optical surface specification
data make inapplicable the simplified formulation of
the PFA commonly used in the literature. We have also derived the
expressions for the Casimir force applicable in the presence
of bubbles or pits. It was shown that surface imperfections
may lead to both a decrease and increase in the magnitude
of the Casimir force up to a few tens of percent
in the range of separations
from 1 to $3\,\mu$m. Keeping in mind that experimentally
it is impossible to determine
the position of points of closest approach between
interacting surfaces with sufficient precision, this in fact
renders lenses of centimeter-size curvature radii
impractical for measurements of the Casimir force.
There might be additional problems in the application of
spherical lenses of centimeter-size curvature radii
for measuring the Casimir force which are not discussed
here. We assert, however, that until the problem of
deviations of lens surface from perfect spherical shape
is somehow resolved any further measurements of the
Casimir force using large spherical lenses are meaningless.
Thus, the single remaining candidate for the registration
of thermal effects in the Casimir force at micrometer
separations is the configuration of two parallel plates.

The fundamental flaw discussed in this paper is not peculiar for
spheres with much smaller radii used in numerous experiments
performed by different authors
by means of an atomic force microscope and micromachined
oscillator (see review \cite{2}).
For instance, surfaces of polystyrene spheres of about
$100\,\mu$m radii made from liquid phase preserve perfect
spherical shape due to surface tension. The surface
quality of such spheres was investigated using a scanning
electron microscope and did not reveal any bubbles and
scratches.

To conclude on a positive note, we agree with Ref.~\cite{1}
that it is the reader who will finally decide which experiments in
the Casimir force area are credible and which are not
``based on verifiable facts and independent scientific
analysis.''

\ack{G.L.K.\ and V.M.M.\ are grateful to the Federal University of
Para\'{\i}ba (Jo\~{a}o Pessoa, Brazil) for kind hospitality.
They were partially supported by CNPq (Brazil).
}

\end{document}